\documentclass[conference]{IEEEtran}
\IEEEoverridecommandlockouts
\usepackage{cite}
\usepackage{amsmath,amssymb,amsfonts}
\usepackage{algorithmic}
\usepackage{graphicx}
\usepackage{textcomp}
\usepackage{xcolor}
\usepackage{colortbl}
\newcommand{\bestcell}[1]{\cellcolor{gray!20}#1}
\def\BibTeX{{\rm B\kern-.05em{\sc i\kern-.025em b}\kern-.08em
    T\kern-.1667em\lower.7ex\hbox{E}\kern-.125emX}}
\begin{document}

\title{TGPP: Trajectory-Guided Plug-and-Play Priors for Sparse Radio Map Reconstruction}

\author{\IEEEauthorblockN{
Jiawen Zhang\IEEEauthorrefmark{1}, 
Zhiyuan Jiang\IEEEauthorrefmark{2},
Sheng Zhou\IEEEauthorrefmark{1}, 
Zhisheng Niu\IEEEauthorrefmark{1}}

\IEEEauthorblockA{\IEEEauthorrefmark{1}Department of Electronic Engineering, Tsinghua University, Beijing, China\\
Email: \{jw-zhang24\}@mails.tsinghua.edu.cn, \{sheng.zhou, niuzhs\}@tsinghua.edu.cn}
\IEEEauthorblockA{\IEEEauthorrefmark{2}School of Communication and Information Engineering, Shanghai University, Shanghai, China\\
Email: jiangzhiyuan@shu.edu.cn}

}

\maketitle
\begin{abstract}
Radio map (RM) reconstruction is essential for environment-aware wireless networks, but practical measurements are often collected along mobility trajectories rather than randomly scattered over the target region. Such trajectory-sampled observations induce spatially heterogeneous uncertainty: near-trajectory regions are directly constrained, whereas distant or occluded regions remain weakly observed, leading to degraded reconstruction accuracy in under-constrained areas. To address this problem, we propose Trajectory-Guided Plug-and-Play Priors (TGPP), a general guidance module for sparse RM reconstruction. TGPP learns an explicit guidance map as an interpretable input-space risk prior, and an implicit guide feature that is projected and fused with backbone hidden representations. TGPP can be attached to different reconstruction backbones without changing their original task formulation. We further introduce RadioFlow-LDM, a latent flow-based generative backbone, and apply TGPP to deterministic, adversarial, graph-based, and latent generative reconstruction models. Experiments on RadioMapSeer with five trajectory sampling rates show that trajectory-sampled reconstruction differs substantially from random sparse interpolation. TGPP improves most reconstruction metrics across backbones, achieving up to $43.1\%$ NMSE reduction relative to the corresponding base backbone without trajectory-guided priors.
\end{abstract}

\begin{IEEEkeywords}
Radio map reconstruction, channel knowledge map, trajectory sampling, trajectory-guided prior, sparse reconstruction, latent flow matching.
\end{IEEEkeywords}

\section{Introduction}
Radio maps (RMs) have become important spatial priors for spectrum awareness, interference management, localization, resource allocation, and trajectory planning \cite{yilmaz2013rem}, \cite{perez2015rem}, \cite{bi2019engineering}. Their broader form, channel knowledge maps (CKMs), further supports environment-aware sixth-generation (6G) communications by encoding location-dependent channel knowledge \cite{zeng2021toward}, \cite{zeng2024ckm}, \cite{wang2026rmtutorial}. However, building high-fidelity RMs from sparse measurements remains challenging due to the high cost of dense measurement campaigns and the computational expense of large-scale ray-tracing simulation \cite{zeng2024ckm}.

\begin{figure}[t]
    \centering
    \setlength{\tabcolsep}{1pt}
    \renewcommand{\arraystretch}{0.9}
    \newcommand{\samplingpanel}[3]{%
        \begin{minipage}[t]{0.46\columnwidth}
            \centering
            \includegraphics[width=\linewidth]{#2.pdf}\\[-0.5mm]
            {\scriptsize #1}
        \end{minipage}%
    }
    \begin{tabular}{cc}
        \samplingpanel{(a) Random/crowdsourced observations}{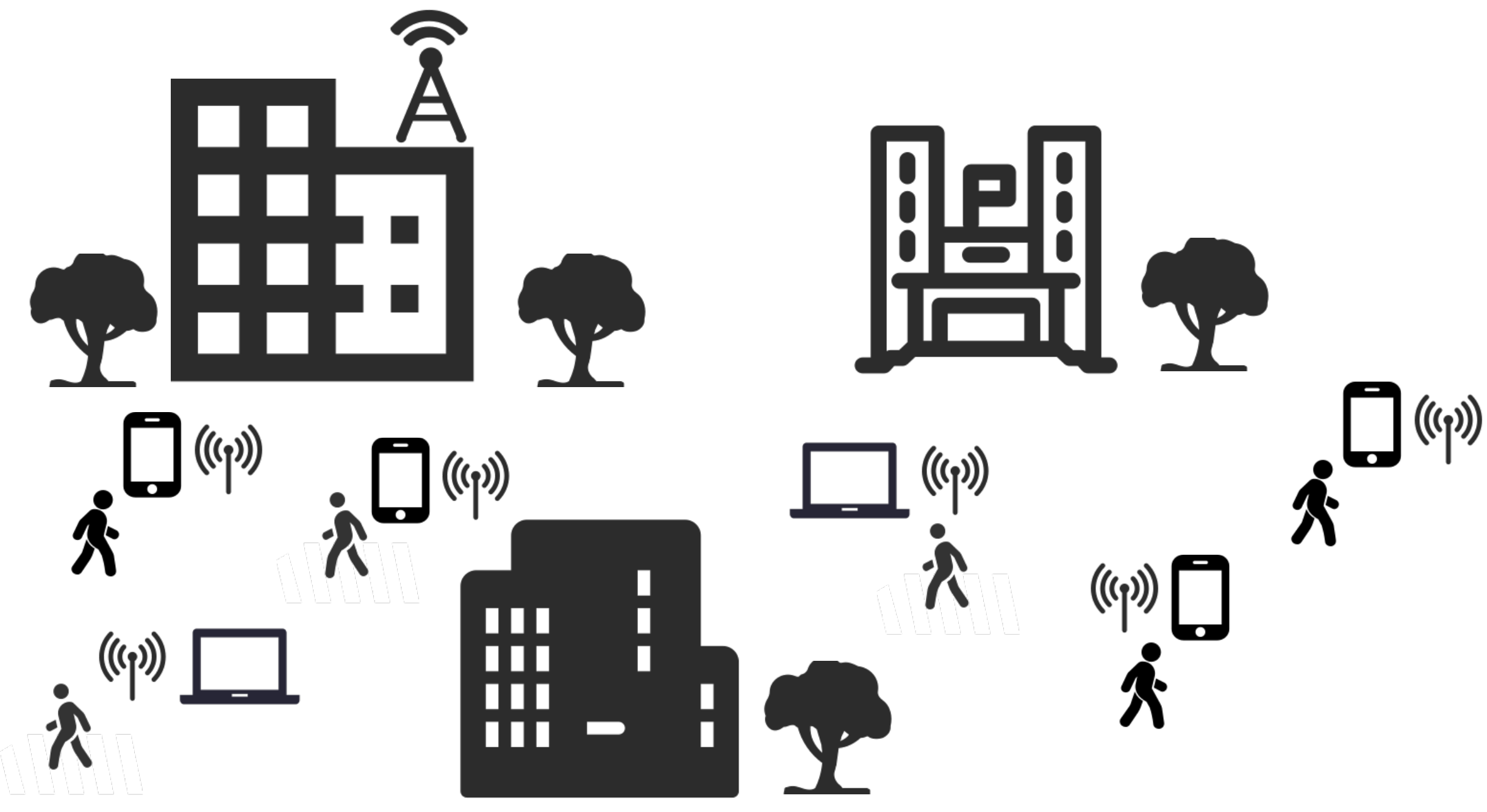}{Random/crowdsourced observations} &
        \samplingpanel{(b) Trajectory-based observations}{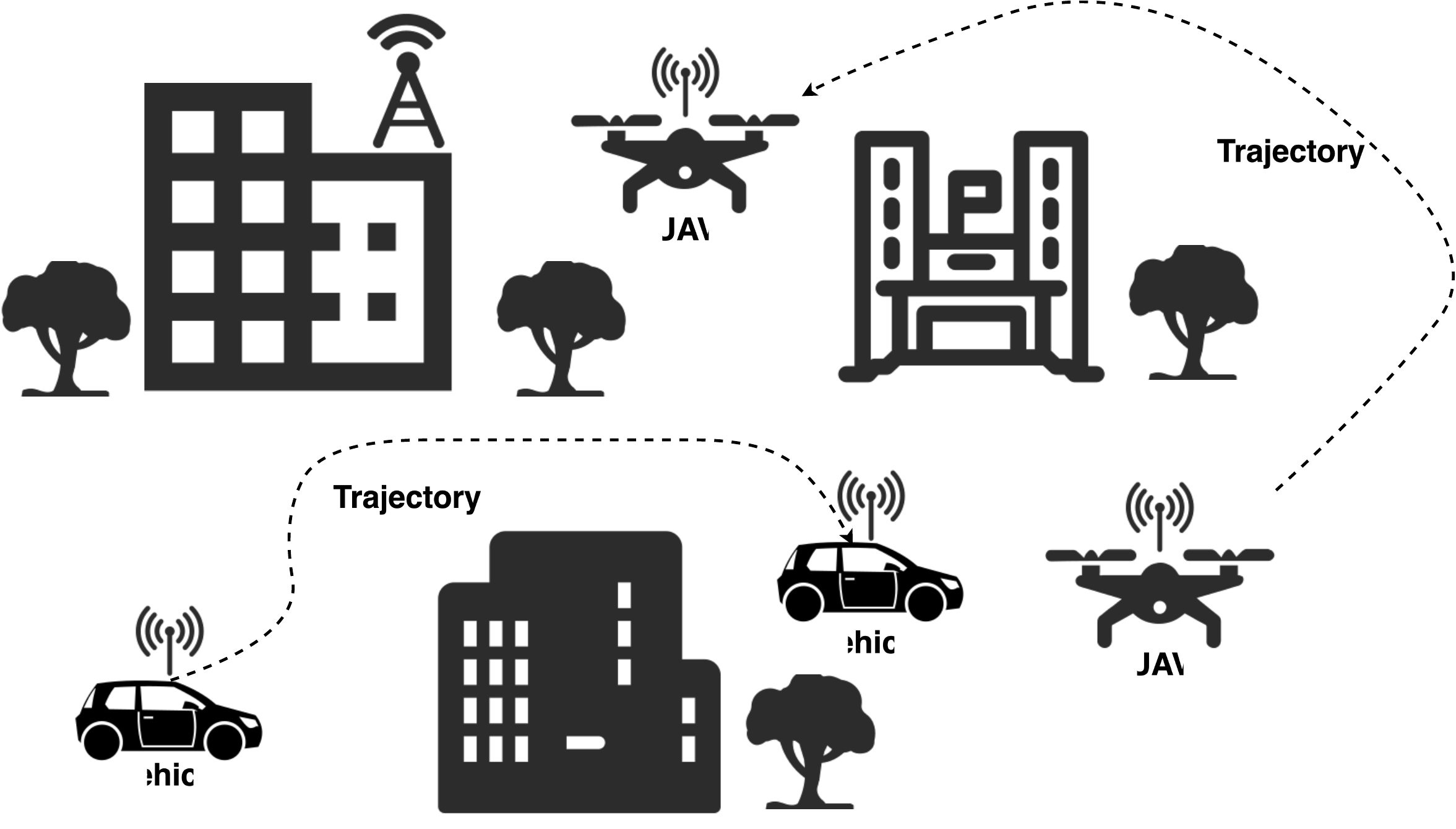}{Trajectory-based observations} \\
        \samplingpanel{(c) Random sampling on RM}{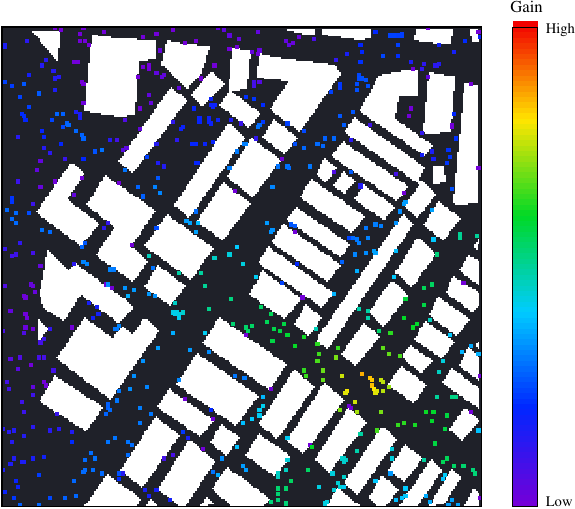}{Random sampling on RM} &
        \samplingpanel{(d) Trajectory sampling on RM}{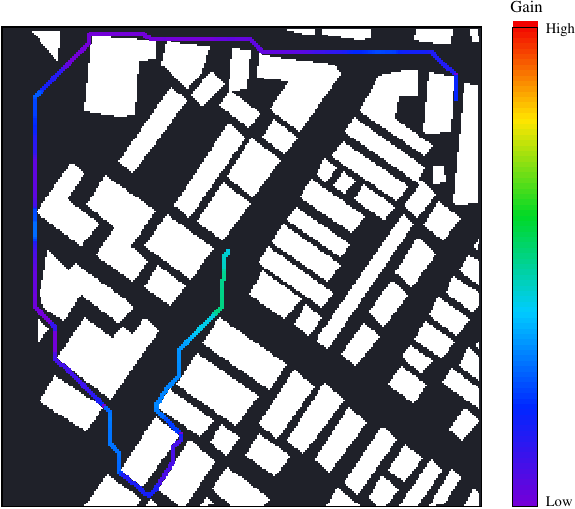}{Trajectory sampling on RM}
    \end{tabular}
    \caption{Comparison between random-sampled and trajectory-sampled observations.}
    \label{fig:random_vs_traj}
\end{figure}

More importantly, practical measurements are often collected along mobility trajectories, such as UAV flight paths, vehicle routes, robot corridors, or user traces \cite{qiu2024uav}, \cite{hu2023trajectorygan}. As illustrated in Fig.~\ref{fig:random_vs_traj}, random and trajectory-based observations differ in both acquisition mechanisms and induced RM constraints: random or crowdsourced observations are spatially scattered, whereas trajectory-sampled observations are concentrated along continuous paths. Consequently, regions near the trajectory are well constrained, while off-trajectory, boundary-adjacent, or occluded regions remain weakly observed. Thus, trajectory-sampled RM reconstruction should be treated as a structured completion problem rather than a conventional random sparse interpolation task.

Existing RM construction methods include interpolation and matrix/tensor completion \cite{sato2017kriging}, \cite{schaufele2019tensor}, CNN-based regression such as RadioUNet \cite{levie2021radiounet}, completion autoencoders \cite{teganya2021autoencoder}, adversarial reconstruction such as RME-GAN \cite{zhang2023rmegan}, graph-based models such as RadioGAT \cite{li2024radiogat}, and recent diffusion, physics-aware, and neural-field methods \cite{wang2024radiodiff}, \cite{wang2025radiodiffk2}, \cite{zhao2023nerf2}. Yet most methods treat sparse observations as ordinary pixels, masks, or graph nodes, without explicitly modeling trajectory-induced uncertainty or injecting trajectory geometry into hidden representations. As a result, \emph{the observation bias induced by trajectory sampling is not directly translated into reconstruction priors}, which limits performance in under-observed and structurally complex regions.

We propose \emph{Trajectory-Guided Plug-and-Play Priors} (TGPP), a general module for trajectory-sampled RM reconstruction. TGPP learns an explicit \emph{guidance map} as an interpretable reconstruction-risk prior and an implicit \emph{guide feature} that is projected and fused into backbone hidden representations. In this way, TGPP provides both input-space risk awareness and representation-space guidance to improve reconstruction in under-observed regions. We further introduce \emph{RadioFlow-LDM}, a latent flow-based generative backbone, and instantiate TGPP on RadioUNet, RME-GAN, RadioGAT, and RadioFlow-LDM to demonstrate plug-and-play adaptability across deterministic, adversarial, graph-based, and latent generative paradigms.

The main contributions are summarized as follows:
\begin{itemize}
\item We formulate trajectory-conditioned RM reconstruction as a structured sparse completion problem with \emph{trajectory-induced spatial bias}.
\item We propose TGPP, which jointly learns an explicit guidance map for input-space risk awareness and an implicit guide feature for hidden-space feature fusion.
\item We introduce RadioFlow-LDM, a latent flow-based generative backbone for compact sparse RM reconstruction, and apply TGPP to RadioUNet, RME-GAN, RadioGAT, and RadioFlow-LDM to demonstrate plug-and-play adaptability across different reconstruction paradigms.
\end{itemize}

The remainder of this paper is organized as follows. Section II presents the system model. Section III introduces TGPP. Section IV describes TGPP-enhanced backbones. Section V reports the experimental setup and evaluation results. Section VI concludes the paper.

\section{System Model and Problem Formulation}
\label{sec:system_model}
We consider single-transmitter RM reconstruction over a regular $H\times W$ grid. Each scene is described by a target map $X\in\mathbb{R}^{1\times H\times W}$ together with a building map $B\in\{0,1\}^{1\times H\times W}$ and a transmitter map $T_x\in\{0,1\}^{1\times H\times W}$. For grid location $\mathbf{p}_{i,j}$, the received signal and average received signal strength (RSS) are
\begin{equation}
    \begin{aligned}
    y(\mathbf{p}_{i,j})&=\sqrt{P_t h(\mathbf{p}_{i,j})}\,s+z(\mathbf{p}_{i,j}),\\
    \Psi(\mathbf{p}_{i,j})&=\mathbb{E}[|y(\mathbf{p}_{i,j})|^2]=P_t h(\mathbf{p}_{i,j})+\sigma^2,
    \end{aligned}
    \label{eq:rx_signal}
\end{equation}
where $h(\mathbf{p}_{i,j})$ is the large-scale channel coefficient determined by transmitter position and environment. In this paper, $X$ denotes the normalized RSS, pathloss, or gain map, depending on the specific dataset construction.

\begin{figure}[t]
    \centering
    \includegraphics[width=0.92\columnwidth,keepaspectratio]{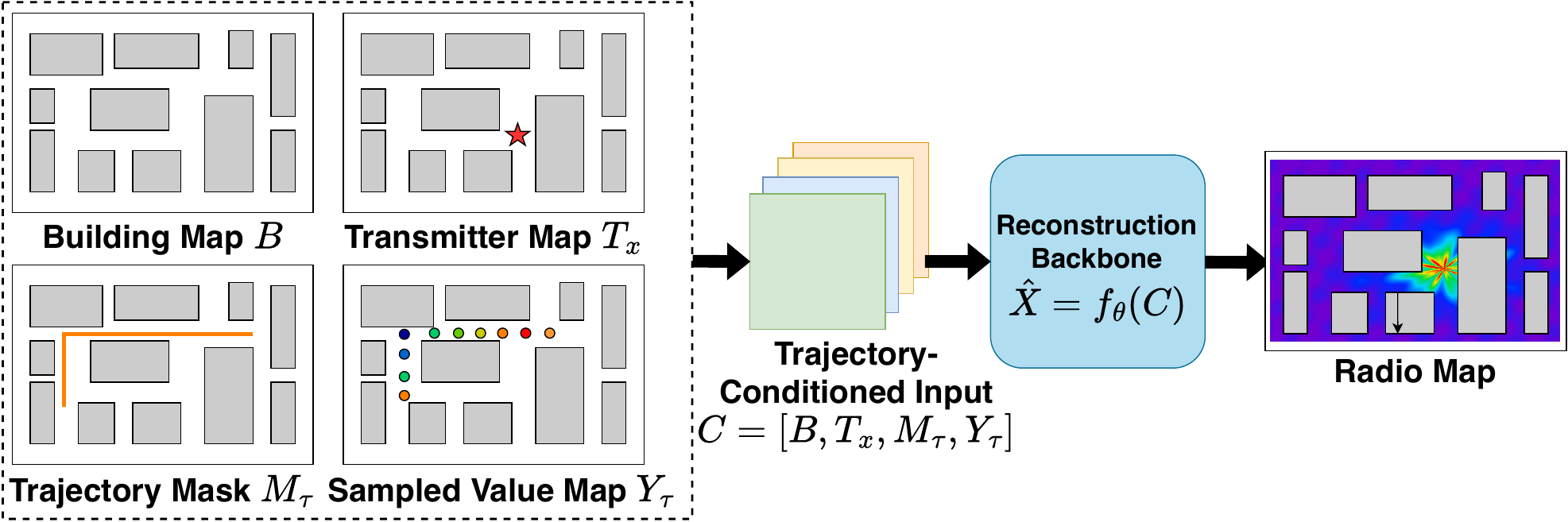}
    \caption{Schematic of trajectory-sampled RM reconstruction.}
    \label{fig:system_model}
\end{figure}

Let $M_\tau$ be the trajectory mask and $Y_\tau$ be the sampled value map, with $Y_\tau=M_\tau\odot X$ in the idealized grid formulation. The trajectory-conditioned input and reconstruction mapping are
\begin{equation}
    C=[B,\;T_x,\;M_{\tau},\;Y_{\tau}],\qquad
    \hat{X}=f_{\theta}(C).
    \label{eq:condition}
\end{equation}
Known samples can be optionally preserved by an observation hard constraint:
\begin{equation}
    \tilde{X}=M_{\tau}\odot Y_{\tau}+(1-M_{\tau})\odot\hat{X}.
    \label{eq:hard_constraint}
\end{equation}
The input-output relationship is summarized in Fig.~\ref{fig:system_model}.

The basic reconstruction objective is
\begin{equation}
    \min_{f_{\theta}}\;\|\hat{X}-X\|_{2}^{2}
    =
    \min_{f_{\theta}}\;\sum_{i,j}|X(i,j)-\hat{X}(i,j)|^{2}.
    \label{eq:basic_objective}
\end{equation}
Trajectory sampling induces spatially heterogeneous supervision: near-trajectory regions are directly constrained, whereas distant or blocked regions must be inferred from environmental structure and contextual cues. This observation motivates the trajectory-guided priors introduced in the next section.

\section{Trajectory-Guided Plug-and-Play Priors}
\label{sec:tgpp}

\subsection{TGPP Overview and Guidance Generator}
TGPP is a trajectory-aware prior module that can be attached to different RM reconstruction backbones without changing their original reconstruction task. Given $C=[B,T_x,M_\tau,Y_\tau]$, TGPP models the spatially heterogeneous uncertainty induced by trajectory sampling and injects such information through input augmentation and hidden-space feature fusion. It guides the backbone to distinguish well-observed regions from poorly constrained areas, and to rely more on environmental priors such as buildings, blockage, and transmitter context in uncertain regions.

TGPP follows a mixed explicit--implicit design. The explicit component is a guidance map that highlights under-observed and structurally challenging regions in the input space. The implicit component is a guide feature that is projected to backbone-specific feature spaces and fused with hidden representations as representation-space guidance.

\begin{figure}[t]
    \centering
    \includegraphics[width=1\columnwidth]{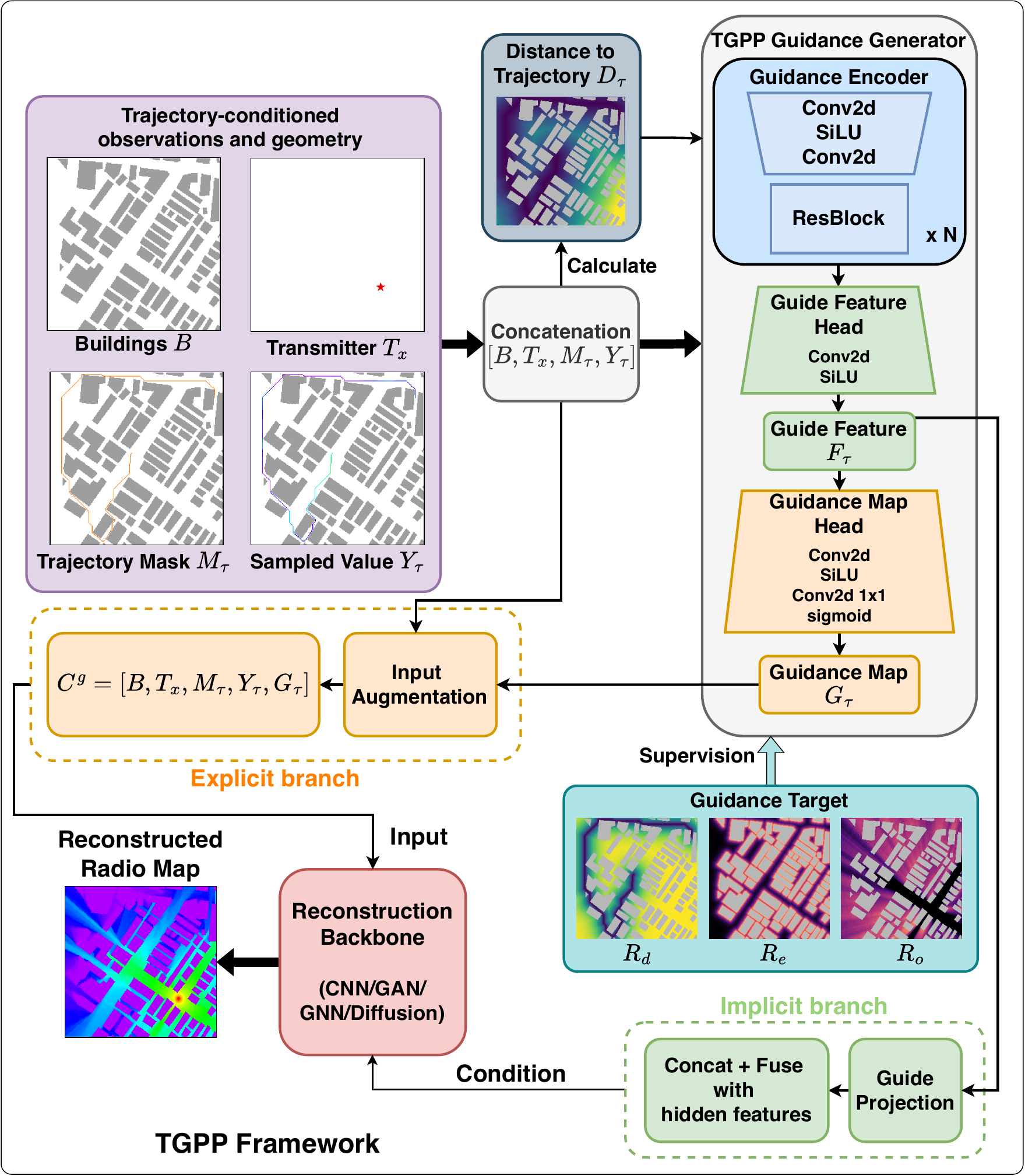}
    \caption{Overview of TGPP. The guidance map augments the model input, while the guide feature is projected and fused with hidden representations of the reconstruction backbone.}
    \label{fig:tgpp_overview}
\end{figure}

To encode trajectory geometry, TGPP uses a distance-to-trajectory map $D_\tau\in\mathbb{R}^{1\times H\times W}$ derived from the trajectory mask. For each grid location, $D_\tau$ records the distance to the nearest sampled trajectory point and is normalized before guidance generation. Unlike the common backbone input in \eqref{eq:condition}, $D_\tau$ is used only by the guidance generator:
\begin{equation}
    (G_\tau,F_\tau)=\mathcal{G}_{\phi}(C,D_\tau),
    \label{eq:guidance_generator}
\end{equation}
where $G_\tau\in\mathbb{R}^{1\times H\times W}$ is the network-predicted explicit guidance map and $F_\tau\in\mathbb{R}^{C_f\times H_f\times W_f}$ is the implicit guide feature, whose channel number and spatial resolution are adapted to the selected backbone. The guidance generator learns trajectory-aware priors by jointly exploiting the environmental layout, transmitter position, trajectory support, sampled values, and distance-to-trajectory geometry.

\subsection{Explicit Guidance Map and Risk Supervision}
The explicit guidance map $G_\tau$ serves as an input-space reconstruction-risk prior and augments the original input as
\begin{equation}
    C^{g}=[B,\;T_x,\;M_\tau,\;Y_\tau,\;G_\tau].
    \label{eq:explicit_guidance_input}
\end{equation}
Unlike a binary observation mask, $G_\tau$ encodes spatially varying reconstruction uncertainty, including off-trajectory, boundary-adjacent, and occlusion-prone regions.

To supervise $G_\tau$, we construct a soft target $R_\tau$ that approximates trajectory-induced reconstruction risk rather than a directly observed physical quantity. It combines three factors: trajectory distance, boundary proximity, and transmitter-to-location occlusion. The distance risk is
\begin{equation}
    R_d(i,j)=1-\exp\!\left(-\frac{d_\tau(i,j)}{\sigma_d}\right),
    \label{eq:distance_risk}
\end{equation}
where $d_\tau(i,j)$ is the distance from $(i,j)$ to the nearest trajectory point. The boundary risk is
\begin{equation}
    R_e(i,j)=\exp\!\left(-\frac{d_e(i,j)}{\sigma_e}\right),
    \label{eq:boundary_risk}
\end{equation}
where $d_e(i,j)$ is the distance to the nearest building boundary extracted from $B$. The occlusion risk approximates transmitter-to-location blockage:
\begin{equation}
    R_o(i,j)=(1-B(i,j))\frac{1}{N_o}\sum_{n=1}^{N_o}
    B\!\left(\ell_n(\mathbf{p}_{\mathrm{tx}},\mathbf{p}_{i,j})\right),
    \label{eq:occlusion_risk}
\end{equation}
where $\ell_n(\mathbf{p}_{\mathrm{tx}},\mathbf{p}_{i,j})$ denotes sampled points along the line segment from the transmitter to $\mathbf{p}_{i,j}$. The non-building mask $(1-B)$ suppresses risk values inside building regions.

The final guidance target is obtained by weighted fusion and smoothing:
\begin{equation}
    \bar{R}_\tau=
    \mathrm{clip}(w_dR_d+w_eR_e+w_oR_o,0,1)\odot(1-B),
    \label{eq:guide_target_raw}
\end{equation}
\begin{equation}
    R_\tau=\mathrm{clip}\!\left(\mathcal{S}_{\sigma_s}(\bar{R}_\tau),0,1\right)\odot(1-B),
    \label{eq:guide_target}
\end{equation}
where $\mathcal{S}_{\sigma_s}(\cdot)$ is a Gaussian smoothing operator. The risk-fusion hyperparameters are specified in Section~\ref{sec:experiments}. The guidance supervision is
\begin{equation}
    \mathcal{L}_{\mathrm{guide}}=\|G_\tau-R_\tau\|_1,
    \label{eq:guide_loss}
\end{equation}
which encourages $G_\tau$ to learn an interpretable risk prior instead of reproducing the sparse observation mask. Since $R_\tau$ is masked by $(1-B)$, the supervision is effectively imposed on non-building regions.

\subsection{Implicit Guide Feature and Hidden-Space Fusion}
While $G_\tau$ provides an explicit input-space prior, $F_\tau$ provides implicit representation-space guidance. For a hidden feature $H_l$ at layer or block $l$, TGPP first projects $F_\tau$ into a scale-aligned guide condition:
\begin{equation}
    Q_l=P_l(F_\tau),
    \label{eq:guide_projection}
\end{equation}
where $P_l(\cdot)$ aligns the guide feature with the spatial or structural resolution and channel dimension of the corresponding backbone representation. The projected guide condition is then concatenated with the backbone feature and fused through a lightweight fusion operator:
\begin{equation}
    \tilde{H}_l
    =
    \mathcal{F}_l\!\left([H_l,\;Q_l]\right),
    \label{eq:feature_fusion}
\end{equation}
where $[\cdot,\cdot]$ denotes feature concatenation and $\mathcal{F}_l(\cdot)$ is a backbone-specific fusion block, such as a convolutional layer, graph projection, or latent feature fuser. This hidden-space fusion allows trajectory-induced difficulty to affect feature extraction, contextual aggregation, and generative refinement inside the backbone.

\subsection{Plug-and-Play Integration and Training Loss}
TGPP is integrated into different reconstruction backbones through two common operations: input augmentation with $G_\tau$ and hidden-space fusion with $F_\tau$. Although the concrete fusion location depends on the backbone architecture, TGPP consistently provides trajectory-derived priors in both the input and representation spaces.

Building on \eqref{eq:basic_objective}, TGPP-enhanced backbones are trained with reconstruction, observation-consistency, guidance-supervision, and task-specific losses:
\begin{equation}
    \mathcal{L}
    =
    \mathcal{L}_{\mathrm{rec}}
    +\lambda_{\mathrm{obs}}\mathcal{L}_{\mathrm{obs}}
    +\lambda_{\mathrm{g}}\mathcal{L}_{\mathrm{guide}}
    +\lambda_{\mathrm{task}}\mathcal{L}_{\mathrm{task}},
    \label{eq:tgpp_objective}
\end{equation}
where $\mathcal{L}_{\mathrm{rec}}$ measures the global discrepancy between $\hat{X}$ and $X$, $\mathcal{L}_{\mathrm{guide}}$ is defined in \eqref{eq:guide_loss}, and
\begin{equation}
    \mathcal{L}_{\mathrm{obs}}=\|M_\tau\odot(\hat{X}-Y_\tau)\|_1.
    \label{eq:obs_loss}
\end{equation}
The observation-consistency loss is applied to the raw reconstruction $\hat{X}$ before optional hard-constraint post-processing, encouraging the backbone itself to preserve known trajectory samples. The task-specific term $\mathcal{L}_{\mathrm{task}}$ represents objectives required by particular backbones, such as adversarial learning for GAN-based reconstruction or latent flow matching for RadioFlow-LDM. Thus, TGPP remains independent of a specific backbone while preserving compatibility with heterogeneous training objectives. The corresponding backbone-specific instantiations are presented in the next section.

\section{TGPP-Enhanced Reconstruction Backbones}
\label{sec:backbones}
The unified TGPP formulation in Section~\ref{sec:tgpp} can be instantiated on different RM reconstruction backbones with minimal architectural modifications. In all TGPP-enhanced variants, the explicit guidance map $G_\tau$ augments the original input as $C^g$, while the implicit guide feature $F_\tau$ is projected and fused with backbone hidden states according to the general form in \eqref{eq:feature_fusion}. Thus, the common design principle is preserved across different reconstruction paradigms, while only the fusion location and the backbone-specific training loss vary with the selected architecture.

\subsection{TGPP-RadioUNet}
RadioUNet formulates RM reconstruction as deterministic image-to-image regression with multi-scale encoder-decoder processing. In TGPP-RadioUNet, $G_\tau$ is concatenated with the input channels, while $F_\tau$ is projected and fused with encoder-decoder hidden states. A typical hidden-state fusion at scale $l$ is
\begin{equation}
    \tilde{H}_l
    =
    \mathrm{Conv}_l\!\left([H_l,\;P_l(F_\tau)]\right),
    \label{eq:tgunet_fusion}
\end{equation}
where $\mathrm{Conv}_l(\cdot)$ denotes a lightweight convolutional fusion block. This allows trajectory-derived priors to affect both local detail recovery and global context aggregation. Training follows \eqref{eq:tgpp_objective}, where $\mathcal{L}_{\mathrm{task}}$ is absorbed into the reconstruction objective.

\subsection{TGPP-RME-GAN}
RME-GAN formulates sparse RM reconstruction as conditional adversarial learning. In TGPP-RME-GAN, TGPP is applied only to the generator: $G_\tau$ augments the generator input and $F_\tau$ is fused with generator hidden states, while the discriminator remains conditioned on the original input $C$.

Let $\hat{X}=G_\theta(C^g \mid F_\tau)$ denote the generated RM, and let $D_\psi(X,C)$ denote the conditional discriminator. We adopt least-squares adversarial learning, where the discriminator loss is
\begin{equation}
    \mathcal{L}_{D}
    =
    \frac{1}{2}\mathbb{E}[(D_\psi(X,C)-1)^2]
    +
    \frac{1}{2}\mathbb{E}[D_\psi(\hat{X},C)^2],
    \label{eq:rmegan_d_loss}
\end{equation}
and the generator is trained with
\begin{equation}
    \mathcal{L}_{G}
    =
    \lambda_{\mathrm{rec}}\mathcal{L}_{\mathrm{rec}}
    +\lambda_{\mathrm{obs}}\mathcal{L}_{\mathrm{obs}}
    +\lambda_{\mathrm{g}}\mathcal{L}_{\mathrm{guide}}
    +\lambda_{\mathrm{adv}}\mathcal{L}_{\mathrm{adv}},
    \label{eq:rmegan_g_loss}
\end{equation}
where $\mathcal{L}_{\mathrm{adv}}=\mathbb{E}[(D_\psi(\hat{X},C)-1)^2]$. The adversarial term encourages conditional realism, while the other losses preserve numerical fidelity and trajectory consistency.

\subsection{TGPP-RadioGAT}
RadioGAT performs patch-level graph modeling, where map patches are represented as graph nodes and updated by graph attention before CNN decoding. In TGPP-RadioGAT, $G_\tau$ is incorporated into the patch-level input representation, while $F_\tau$ is projected to node resolution and fused with graph embeddings:
\begin{equation}
    \tilde{\mathbf{h}}_v
    =
    \mathcal{F}_{\mathrm{gat}}
    \!\left([\mathbf{h}_v,\;P_{\mathrm{gat}}(F_\tau)_v]\right),
    \label{eq:radiogat_fusion}
\end{equation}
where $\mathbf{h}_v$ denotes the embedding of graph node $v$. This fusion enriches node states with trajectory-derived priors. Training follows \eqref{eq:tgpp_objective}, where $\mathcal{L}_{\mathrm{task}}$ corresponds to the graph-based reconstruction objective.

\subsection{RadioFlow-LDM and TGPP-RadioFlow-LDM}
We further introduce RadioFlow-LDM, a latent flow-based generative backbone for sparse RM reconstruction. It first maps radio maps into a compact latent space through an encoder-decoder pair,
\begin{equation}
    z_1=E_\omega(X),\qquad X\approx D_\eta(z_1),\qquad \hat{X}=D_\eta(\hat{z}_1),
    \label{eq:radioflow_vae}
\end{equation}
where $z_1$ denotes the data latent and $\hat{z}_1$ denotes the generated latent. A noise latent $z_0\sim\mathcal{N}(0,I)$ is transported toward the data latent through
\begin{equation}
    z_t=(1-t)z_0+t z_1,\qquad t\in[0,1],
    \label{eq:radioflow_path}
\end{equation}
with latent dynamics
\begin{equation}
    \frac{d z_t}{dt}=v_\theta(z_t,t,C),
    \label{eq:radioflow_ode}
\end{equation}
where $v_\theta$ is conditioned on $C$. The flow-matching objective is
\begin{equation}
    \mathcal{L}_{\mathrm{flow}}
    =
    \mathbb{E}_{t,z_0,z_1}
    \left[
    \left\|
    v_\theta(z_t,t,C)-(z_1-z_0)
    \right\|_2^2
    \right].
    \label{eq:radioflow_loss}
\end{equation}
Latent-space flow matching reduces spatial complexity while retaining generative flexibility. At inference, the learned flow transports $z_0$ to $\hat{z}_1$, which is decoded into the reconstructed RM.

TGPP-RadioFlow-LDM integrates TGPP into both the conditioning input and the latent flow-matching network. The TGPP-enhanced version uses
\begin{equation}
    \begin{aligned}
    C_{\mathrm{ldm}}^g
    &=
    \mathcal{F}_{\mathrm{ldm}}
    \!\left([E_c(C^g),\;P_{\mathrm{ldm}}(F_\tau)]\right),\\
    \frac{d z_t}{dt}
    &=v_\theta(z_t,t,C_{\mathrm{ldm}}^g),
    \end{aligned}
    \label{eq:tgpp_radioflow_ode}
\end{equation}
where $E_c(\cdot)$ encodes $C^g$, $P_{\mathrm{ldm}}(F_\tau)$ provides the latent-scale guide feature, and $\mathcal{F}_{\mathrm{ldm}}(\cdot)$ fuses them into the condition feature of the latent flow U-Net.

Training follows \eqref{eq:tgpp_objective}, with $\mathcal{L}_{\mathrm{task}}=\mathcal{L}_{\mathrm{flow}}$. The effectiveness of these TGPP-enhanced backbones is evaluated in Section~\ref{sec:experiments}.

\section{Experimental Setup and Results}
\label{sec:experiments}

\subsection{Simulation Setup}
We conduct simulations on the RadioMapSeer dataset \cite{yapar2022dataset}, using building maps, transmitter maps, and ray-tracing-based gain maps as the environment-conditioned RM reconstruction data. The map IDs are split into training and test sets following the default partition, where IDs $0$--$499$ are used for training and IDs $500$--$700$ are used for testing. All gain maps are normalized to $[0,1]$.

We evaluate sparse RM reconstruction under both random-sampled and trajectory-sampled observation patterns. Five sampling rates are considered: $0.5\%$, $1.0\%$, $1.5\%$, $2.0\%$, and $2.5\%$. For a sampling rate $r$, the number of observed pixels is set according to the number of non-building accessible pixels. The random-sampled mask uniformly selects accessible pixels. The trajectory-sampled mask is generated by connecting randomly selected accessible waypoints with collision-free A* paths over non-building pixels until the target observation budget is reached. Eight mask variants are generated for each map and sampling rate. The sampled value map is then obtained as $Y_\tau=M_\tau\odot X$, and the distance-to-trajectory map used by TGPP is computed from $M_\tau$.

All models are trained and evaluated under the same data split, sampling rates, and non-building evaluation mask. Unless otherwise stated, the reconstructed output is evaluated after enforcing observation consistency at sampled locations. We report mean absolute error (MAE), root mean squared error (RMSE), normalized mean squared error (NMSE), peak signal-to-noise ratio (PSNR), and structural similarity (SSIM). PSNR is computed with the normalized peak value set to one. For TGPP, the guide target in \eqref{eq:guide_target} uses $\sigma_d=16$, $\sigma_e=5$, $\sigma_s=1$, $N_o=64$, and $(w_d,w_e,w_o)=(0.6,0.25,0.15)$.

\subsection{Results and Discussion}
\begin{figure*}[t]
    \centering
    \setlength{\tabcolsep}{2pt}
    \renewcommand{\arraystretch}{0.9}
    \newcommand{\rmvispanel}[2]{%
        \begin{minipage}[t]{0.19\textwidth}
            \centering
            \includegraphics[width=\linewidth]{#2}\\[-0.5mm]
            {\scriptsize #1}
        \end{minipage}%
    }
    \begin{tabular}{ccccc}
        \rmvispanel{(a) Ground truth}{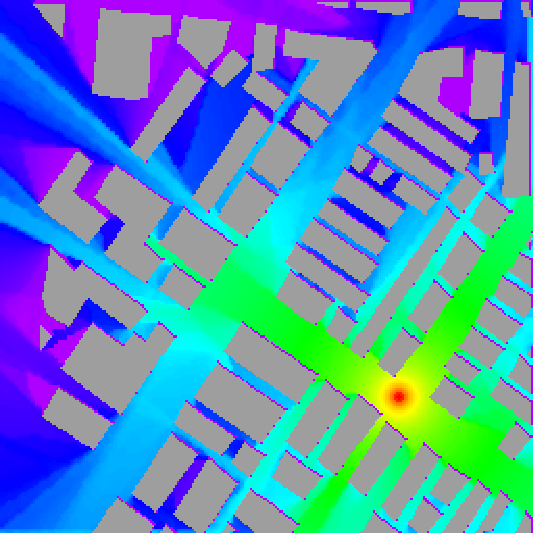} &
        \rmvispanel{(b) RadioUNet}{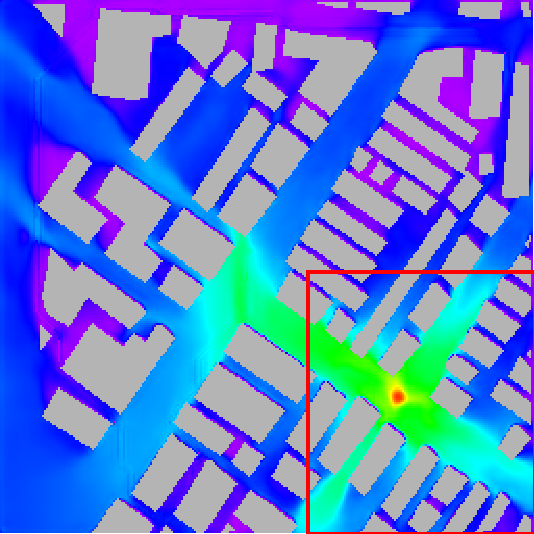} &
        \rmvispanel{(c) RME-GAN}{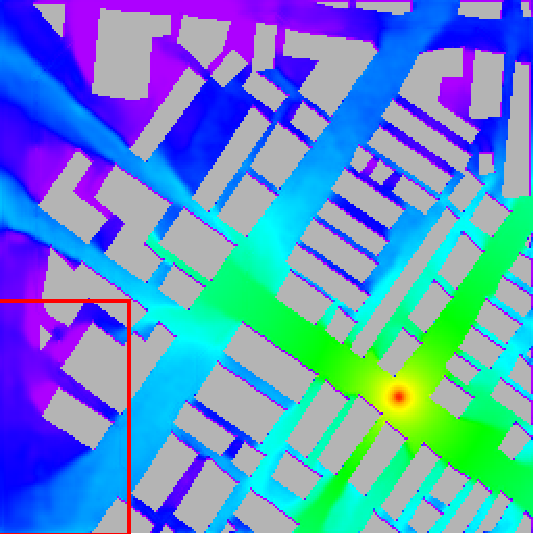} &
        \rmvispanel{(d) RadioGAT}{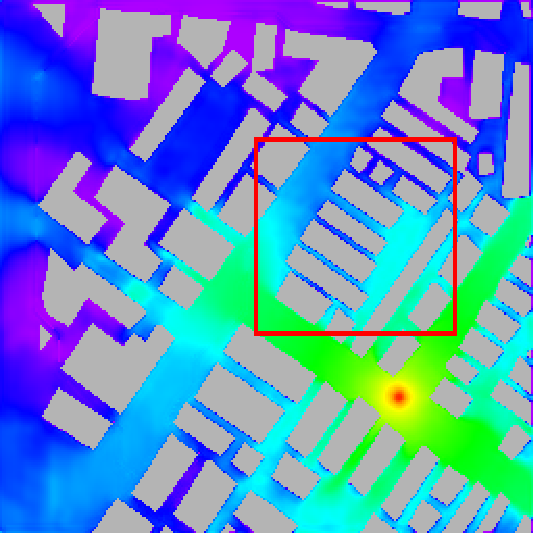} &
        \rmvispanel{(e) RadioFlow-LDM}{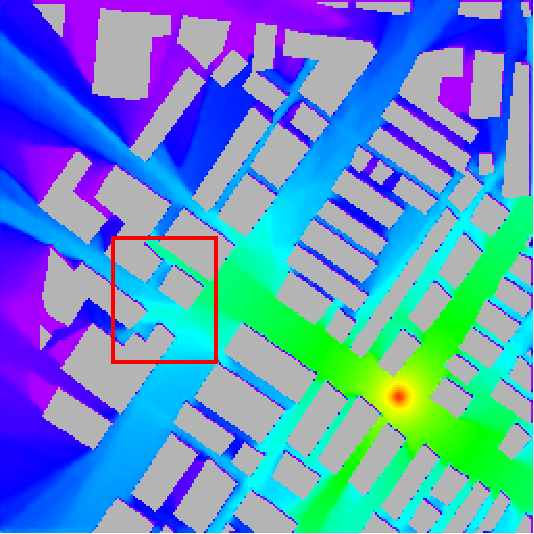} \\
        \rmvispanel{(f) Traj.+building}{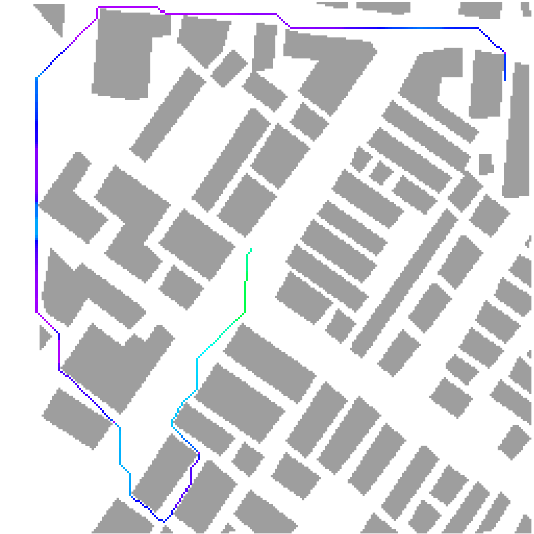} &
        \rmvispanel{(g) TGPP-RadioUNet}{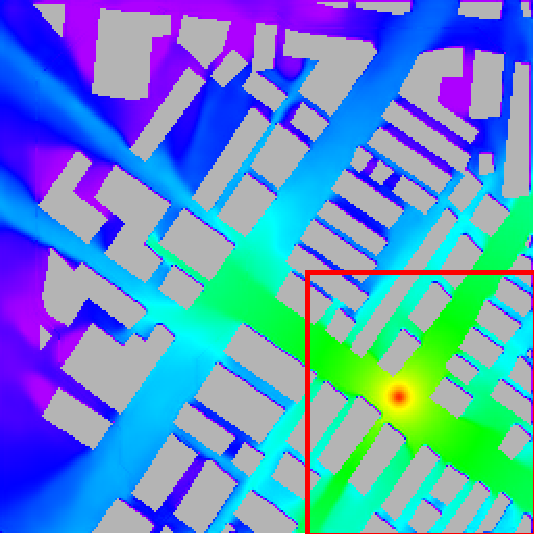} &
        \rmvispanel{(h) TGPP-RME-GAN}{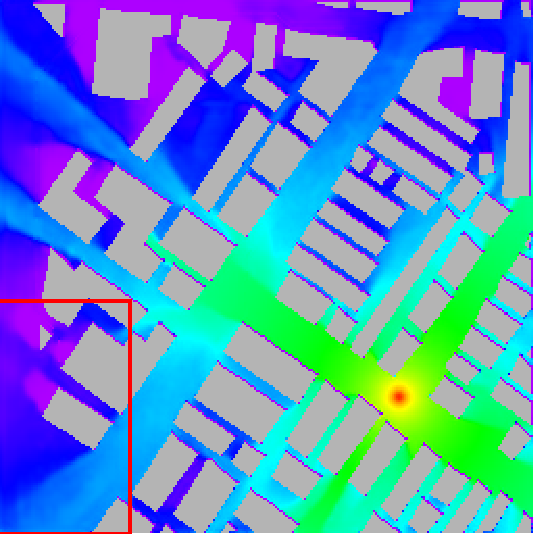} &
        \rmvispanel{(i) TGPP-RadioGAT}{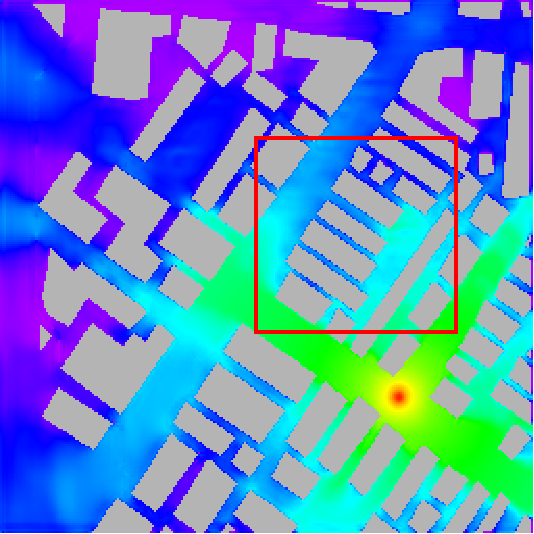} &
        \rmvispanel{(j) TGPP-RadioFlow-LDM}{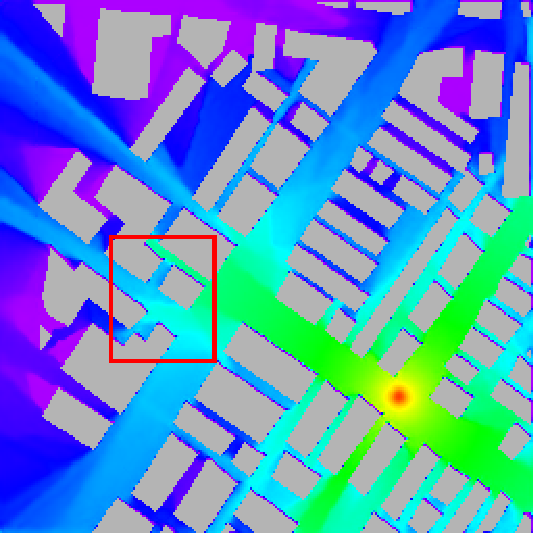}
    \end{tabular}
    \caption{Qualitative comparison of RM reconstruction under trajectory-sampled observations. TGPP-enhanced variants generally produce smoother local structures and more consistent off-trajectory reconstruction than their corresponding base backbones.}
    \label{fig:qualitative_rm}
\end{figure*}

\textbf{Qualitative visualization.}
Fig.~\ref{fig:qualitative_rm} shows a representative trajectory-sampled reconstruction case, where off-trajectory regions must be inferred from the building layout, transmitter context, and neighboring propagation patterns. The red boxes facilitate local comparison between each backbone and its TGPP-enhanced counterpart. TGPP produces smoother and more spatially consistent reconstructions in these regions, indicating that the guidance map and guide feature help exploit trajectory-derived and environmental priors.

\textbf{Cross-domain behavior.}
Table~\ref{tab:cross_domain_radiounet} demonstrates a clear distribution gap between random and trajectory sampling. RadioUNet trained on random samples degrades sharply when tested on trajectory samples, with MAE, RMSE, and NMSE increasing by $7.20\times$, $3.02\times$, and $9.14\times$, respectively, and PSNR dropping by $9.61$ dB. In contrast, trajectory-trained RadioUNet transfers to random samples with only mild degradation. This asymmetric behavior confirms that trajectory-sampled RM reconstruction is a structured and biased completion problem rather than a simple extension of random sparse interpolation.

\textbf{Effect of guidance map and guide feature.}
Table~\ref{tab:tgpp_ablation} verifies that the two TGPP components are complementary. The guidance-map-only variant mainly suppresses large structural errors, reducing RMSE and NMSE by $17.0\%$ and $31.1\%$. The guide-feature-only variant provides stronger overall gains through hidden-space feature fusion, reducing MAE, RMSE, and NMSE by $7.4\%$, $23.2\%$, and $41.0\%$. The full TGPP achieves the best performance on all metrics, showing that input-space risk awareness and representation-space guidance jointly improve trajectory-sampled reconstruction.

\textbf{Overall comparison.}
Table~\ref{tab:overall_comparison} compares each backbone with its TGPP-enhanced variant. TGPP brings the clearest gains to RadioUNet, reducing MAE, RMSE, and NMSE by $9.4\%$, $24.6\%$, and $43.1\%$, and also improves RME-GAN across all metrics. For RadioGAT, whose graph neural network already models relationships among sampled regions and spatial patches, TGPP provides smaller marginal gains but still improves RMSE, NMSE, PSNR, and SSIM. RadioFlow-LDM also benefits moderately, as its latent generative backbone already captures stronger global structure. Overall, TGPP is broadly compatible with heterogeneous reconstruction paradigms, with the largest gains on convolutional and adversarial backbones.

\begin{table}[!t]
\centering
\scriptsize
\setlength{\tabcolsep}{2pt}
\caption{Cross-domain behavior of RadioUNet under different sampling distributions. Results are averaged over five sampling rates: $0.5\%$, $1.0\%$, $1.5\%$, $2.0\%$, and $2.5\%$.}
\label{tab:cross_domain_radiounet}
\begin{tabular}{c c c c c c c}
\hline
Train & Test & MAE$\downarrow$ & RMSE$\downarrow$ & NMSE$\downarrow$ & PSNR$\uparrow$ & SSIM$\uparrow$ \\
\hline
Random & Random & 0.0078 & 0.0285 & 0.0068 & 30.91 & 0.9511 \\
Random & Trajectory & 0.0561 & 0.0861 & 0.0622 & 21.30 & 0.8628 \\
\multicolumn{2}{c}{Relative change} & 7.20$\times$ & 3.02$\times$ & 9.14$\times$ & -9.61 & -0.0882 \\
\hline
Trajectory & Trajectory & 0.0166 & 0.0449 & 0.0169 & 26.95 & 0.9016 \\
Trajectory & Random & 0.0182 & 0.0462 & 0.0179 & 26.71 & 0.8948 \\
\multicolumn{2}{c}{Relative change} & 1.09$\times$ & 1.03$\times$ & 1.06$\times$ & -0.24 & -0.0069 \\
\hline
\end{tabular}
\vspace{1mm}

\caption{Ablation of TGPP components on the RadioUNet backbone. Results are averaged over the five sampling rates.}
\label{tab:tgpp_ablation}
\begin{tabular}{l c c c c c c c}
\hline
Model & Map & Feat. & MAE$\downarrow$ & RMSE$\downarrow$ & NMSE$\downarrow$ & PSNR$\uparrow$ & SSIM$\uparrow$ \\
\hline
RadioUNet & -- & -- & 0.0166 & 0.0449 & 0.0169 & 26.95 & 0.9016 \\
TGPP-RadioUNet-map & $\checkmark$ & -- & 0.0166 & 0.0373 & 0.0117 & 28.56 & 0.9251 \\
TGPP-RadioUNet-feature & -- & $\checkmark$ & 0.0154 & 0.0345 & 0.0100 & 29.24 & 0.9331 \\
TGPP-RadioUNet & $\checkmark$ & $\checkmark$ & 0.0151 & 0.0339 & 0.0096 & 29.40 & 0.9341 \\
\hline
\end{tabular}
\vspace{1mm}

\caption{Overall performance comparison between base backbones and TGPP-enhanced variants. Results are averaged over the five sampling rates.}
\label{tab:overall_comparison}
\begin{tabular}{l c c c c c}
\hline
Model & MAE$\downarrow$ & RMSE$\downarrow$ & NMSE$\downarrow$ & PSNR$\uparrow$ & SSIM$\uparrow$ \\
\hline
RadioUNet & 0.0166 & 0.0449 & 0.0169 & 26.95 & 0.9016 \\
TGPP-RadioUNet & \bestcell{0.0151} & \bestcell{0.0339} & \bestcell{0.0096} & \bestcell{29.40} & \bestcell{0.9341} \\
\hline
RME-GAN & 0.0137 & 0.0307 & 0.0079 & 30.24 & 0.9409 \\
TGPP-RME-GAN & \bestcell{0.0126} & \bestcell{0.0269} & \bestcell{0.0061} & \bestcell{31.39} & \bestcell{0.9497} \\
\hline
RadioGAT & \bestcell{0.0226} & 0.0510 & 0.0218 & 25.85 & 0.8886 \\
TGPP-RadioGAT & 0.0231 & \bestcell{0.0508} & \bestcell{0.0217} & \bestcell{25.88} & \bestcell{0.8897} \\
\hline
RadioFlow-LDM & \bestcell{0.0143} & 0.0295 & 0.0073 & 30.61 & 0.9432 \\
TGPP-RadioFlow-LDM & \bestcell{0.0143} & \bestcell{0.0289} & \bestcell{0.0070} & \bestcell{30.79} & \bestcell{0.9450} \\
\hline
\end{tabular}
\end{table}

\section{Conclusion}
\label{sec:conclusion}
This paper studied sparse RM reconstruction under trajectory-sampled observations, where measurements are concentrated along mobility paths and induce spatially biased reconstruction uncertainty. To address this challenge, we proposed TGPP, a plug-and-play trajectory guidance module that learns an explicit guidance map for input-space risk awareness and an implicit guide feature for hidden-space feature fusion. We further introduced RadioFlow-LDM as a latent flow-based generative backbone and applied TGPP to deterministic, adversarial, graph-based, and latent generative reconstruction models. The cross-domain results show that a RadioUNet trained on random samples suffers a $9.14\times$ NMSE increase when tested on trajectory samples, confirming a clear mismatch between the two sampling distributions. Across backbones, TGPP improves most reconstruction metrics, with up to $43.1\%$ NMSE reduction on RadioUNet and consistent gains on RME-GAN, RadioGAT, and RadioFlow-LDM. Future work will extend TGPP to three-dimensional trajectory sampling and further integrate UAV trajectory planning for efficient radio-map acquisition.

\end{document}